\documentclass[11pt,twoside]{article}
\pdfoutput=1
\usepackage{asp2006}
\usepackage{graphicx}
\usepackage{hyperref}

\begin{document}
\title{A Shared Tully-Fisher Relation for Spiral and S0 Galaxies}
\author{M. J. Williams, M. Bureau, and M. Cappellari}
\affil{Sub-department of Astrophysics, University of Oxford, Denys
Wilkinson Building, Keble Road, Oxford OX1 3RH, United Kingdom}

\begin{abstract} 
We measure the Tully-Fisher relations of 14 lenticular
galaxies (S0s) and 14 spirals. We use two measures of rotational
velocity. One is derived directly from observed spatially-resolved
stellar kinematics and the other from the circular velocities of mass
models that include a dark halo and whose parameters are constrained by
detailed kinematic modelling. Contrary to the naive expectations of
theories of S0 formation, we find no significant difference between the
Tully-Fisher relations of the two samples when plotted as functions of
both brightness and stellar mass.
\end{abstract}

\vspace{-1em}

\section{Introduction}

The Tully-Fisher (TF) relation is a widely-used and strong correlation
between the maximum rotational velocity of spiral galaxies and their
total magnitude. It follows naturally from the assumption that spiral
galaxies have similar surface brightness profiles and approximately
equal dynamical mass-to-light ratios for a given mass.

Many authors have argued that at least some S0 galaxies are the faded
direct descendants of spiral galaxies
\citep[e.g.][]{Dressler:1980,Dressler:1997}. Environmental
processes such as strangulation \citep[e.g.][]{Larson:1980} or ram-pressure
stripping \citep[e.g.][]{Gunn:1972} may have stripped these galaxies of
their gas and left them unable to form stars. This process should make
only a slight change to their dynamical masses, but a significant change
to their luminosities. Regardless of the particular mechanism by which S0s
form, they are therefore generally expected to have higher mass-to-light
ratios than spirals on average. They should therefore have
fainter luminosities for a given rotational velocity and lie below the
TF relation defined by spiral galaxies.

Here we seek to test this prediction using long-slit stellar kinematics
and detailed mass models for a sample of S0 galaxies and a control
sample of spiral galaxies. A secondary goal is to investigate whether
the size of the offset between the S0 and spiral TF relations (if one
exists) changes when we plot a TF-like relation which uses mass
rather than luminosity, as expected from the S0 formation mechanisms
discussed above.

\section{Data and Models}

We use a sample of 14 early spirals (Sa and Sb) and 14 S0s. All of the
galaxies are close to edge-on. Many of them have boxy bulges, which are
believed to be bars viewed side-on
\citep[e.g.][]{Kuijken:1995,Bureau:1999}. The bar should not, however,
affect the present results because we use maximum rotational velocities
well outside of the bar regions and the bar fraction in the sample
($\approx 75$ per cent) is representative of that in the local spiral galaxy
population ($\approx 65$ per cent; see, e.g., \citealt{Sheth:2008} and
references therein).

\begin{figure}
\centering
\includegraphics[width=6cm]{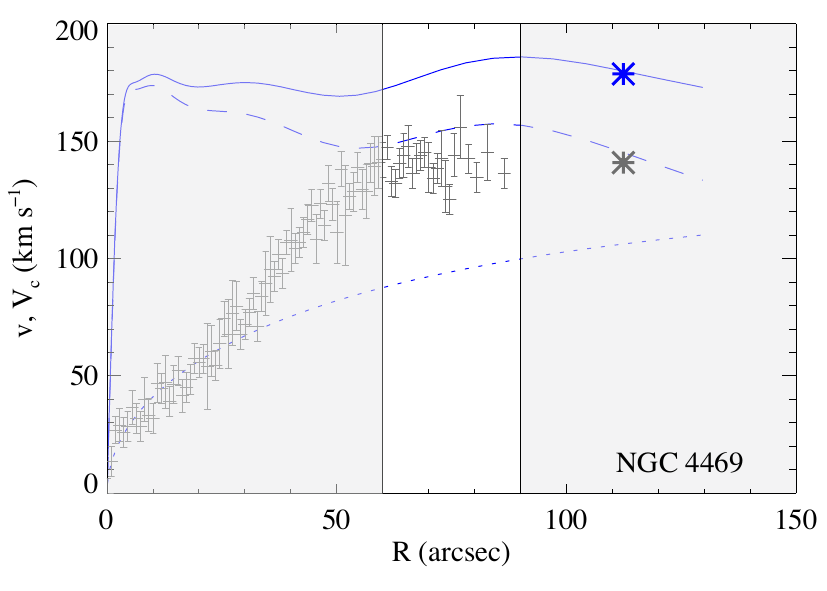}
\includegraphics[width=6cm]{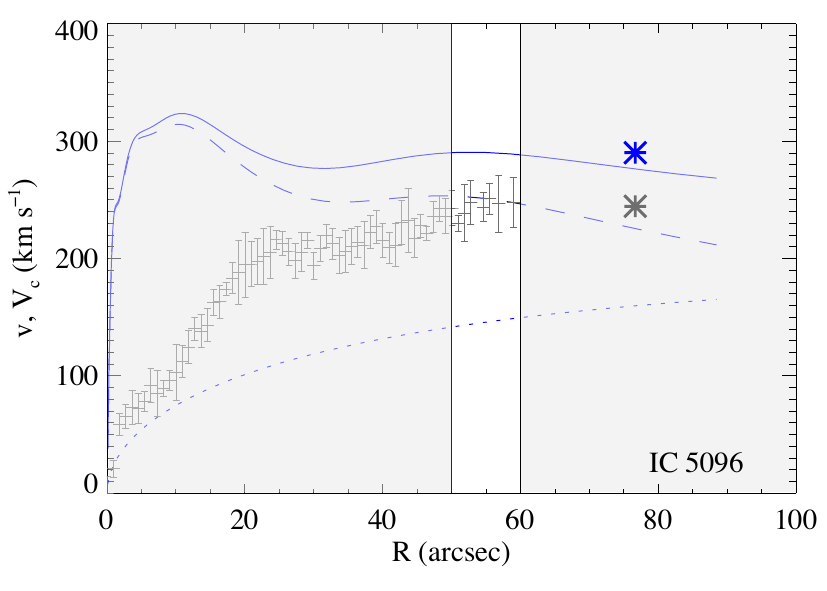}
\caption{Example rotation curves and circular velocity profiles. The
data points show the observed line-of-sight velocities. The lines show
the total circular velocities of the axisymmetric mass models (solid),
the stellar components (dashed line) and dark halo components (dotted
line). The kinematic quantities used in this work, $v$ and $V_c$, are
the means of the data and model in the unshaded radial regions,
chosen to be the flat regions of the observed rotation curves. They are
shown here as large stars. The difference between the rotation curves and
circular velocity profiles is due to line-of-sight integration,
projection effects and pressure support.}
\label{fig:rotcurve}
\end{figure}

The major-axis long-slit stellar kinematic observations are presented in
\cite{Chung:2004}. These data were observed and reduced identically for
the spirals and S0s. We measure the maximum observed rotation velocity
$v$ directly from the stellar kinematics by taking the mean of the data points
in the flat region of the rotation curve. Two sample rotation curves are
shown in Fig.~\ref{fig:rotcurve}.

In \cite{Williams:2009} we modelled the mass distribution of each galaxy
by assuming it is composed of an axisymmetric stellar component with
a constant mass-to-light ratio and a NFW dark halo
\citep{Navarro:1997}. We determined the two parameters of the mass
models (the stellar mass-to-light ratio and total halo mass) by
comparing the observed second velocity moment to that predicted by
solving the Jeans equations assuming a constant anisotropy
\citep{Cappellari:2008}.

We compute the circular velocity curves of the galaxies from these
models. The circular velocity is of course free from the effects of
projection and asymmetric drift, which is particularly important when
comparing S0s and spirals, because S0s have greater pressure support. We
characterize the circular velocity by a single value $V_c$ by taking its
average at the same radii that were used to measure the maximum observed
rotational velocity (see Fig.~\ref{fig:rotcurve}).

We adopt errors in the observed velocity of half the difference between
the approaching and receding sides. The uncertainties in the circular
velocity curves are due to the errors in the parameters of the mass
model, described in \cite{Williams:2009}. 

We use total absolute magnitudes $M_K$ at $K$-band derived from apparent
magnitudes (and errors) taken from the 2MASS Extended Source Catalog
\citep{Skrutskie:2006} and distances (and errors) from the NASA/IPAC
Extragalactic Database. $K$-band is chosen to minimize the effects of
obscuration by dust, which can be significant in edge-on systems at
optical wavelengths. We convert these to stellar masses $M_*$ using
the best-fitting stellar mass-to-light ratios presented in
\cite{Williams:2009}.

\begin{figure}[t]
\centering
\includegraphics{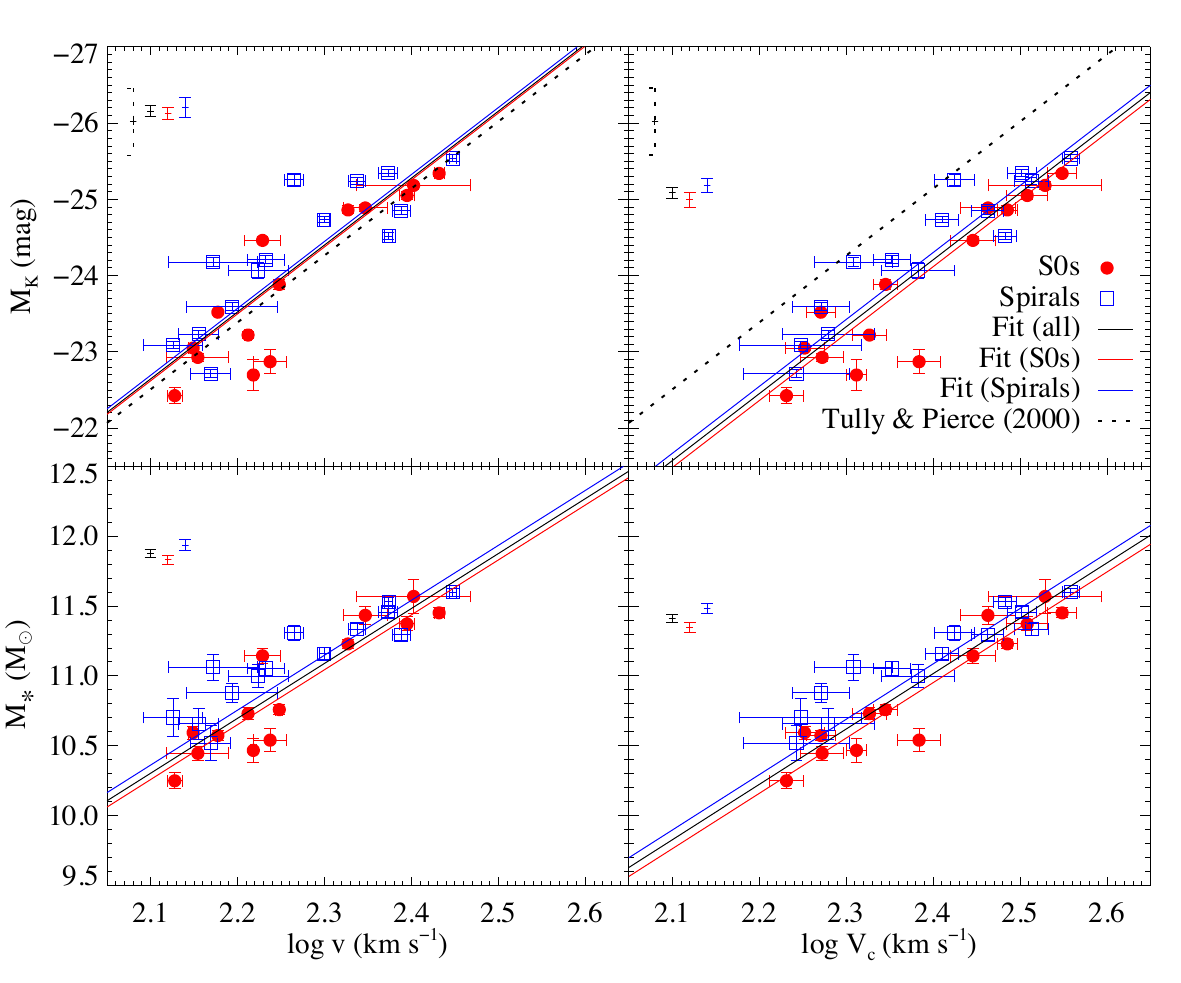}
\caption{TF-like plots for our samples. Spirals are shown as blue squares
and S0s as red triangles. We show power law fits of the form $M_{K} = a
+ b (\log v - 2.5)$ for the two samples as red and blue lines, and
fits to all 28 galaxies as solid black lines. We show $1\,\sigma$
confidence intervals on $a$ for each sample in the top-left of each
plot. The TF relation for spirals found by \protect{\cite{Tully:2000}}
is also shown as a dashed black line in the upper plots.}
\label{fig:tf}
\end{figure}

We present four TF-like plots derived from the above quantities in
Fig.~\ref{fig:tf}. We first plot the maximum observed rotational
velocity and model circular velocity against the total magnitude to
provide both a purely observed TF relation and one which eliminates
asymmetric drift. We also plot both velocities against the stellar mass
of the system, $M_*$.

It is important to note that at no point in this analysis do we do
anything that might systematically affect the S0s in the sample
differently to the spirals.

\section{Discussion}

In neither of the TF-like plots showing measures of velocity against
measures of magnitude (upper panes of Fig.~\ref{fig:tf}) do we see a
significant difference between the spirals and S0s. This is demonstrated
by the error bars in the top-left corner of each plot, which show the
$1\,\sigma$ confidence interval of $a$ in fits to each sample, assumed
to be of the form $M_{K} = a + b (\log v - 2.5)$. We fixed the gradients
of the individual fits in the upper panels at $b = -8.78$, as found by
\cite{Tully:2000}.

When we use circular velocity as the measure of rotation, which avoids
the effects of asymmetric drift (top-right pane of Fig.~\ref{fig:tf}).
Both samples (S0s \emph{and} spirals) are significantly offset
from the TF relation found by \cite{Tully:2000}, which is derived from
unresolved H\textsc{i} kinematics (single-dish spectra).\footnote{We
follow \cite{Bedregal:2006} in shifting the \cite{Tully:2000} $K$-band
TF relation by $-0.207$\,mag to match our adopted $H_0 =
70\mathrm{\,km\,s^{-1}\,Mpc^{-1}}$.} This offset is in the same sense
and of approximately the same size ($\approx +1\,$mag or $+0.1$\,dex in
velocity) as that found by \cite{Bedregal:2006} for their sample of S0s
only.

There is a marginal change in the offset between the two samples in the
change from magnitude to stellar mass (lower panels), although this
could be due to a different slope in the best fits of the two samples.
In this work we fixed the gradients of each fit to $M_*$ to the common
gradient of the two samples (3.92 as a function of $\log v$ and 3.96 for
$\log V_c$).

In conclusion, we find no evidence that S0s lie in a different region of
the luminosity/mass--rotational velocity plane than spirals. Present
models of S0 formation, which our observations seem to contradict, are
nevertheless extremely appealing. Future work will head in two
directions: firstly, we will further attempt to determine the origin of
the offset between our TF relations and that of \cite{Tully:2000} as
well as the difference between our findings and those of
\cite{Bedregal:2006}. Secondly, we will characterize the statistical
significance of the (lack of) offset and the scatter in the TF relations
of our samples at both $K$-band and $B$-band, and seek to interpret the
results within the context of models of galaxy evolution.

\end{document}